\documentclass[12pt]{article}
\pdfoutput=1
\usepackage{bm,amsmath,amssymb,graphicx}

\begin{document}
\pagenumbering{arabic}
\begin{titlepage}

\title{Gravitational waves in massive conformal gravity}

\author{F. F. Faria$\,^{*}$ \\
Centro de Ci\^encias da Natureza, \\
Universidade Estadual do Piau\'i, \\ 
64002-150 Teresina, PI, Brazil}

\date{}
\maketitle

\begin{abstract}
First, we obtain the plane wave solution of the linearized massive conformal 
gravity field equations. It is shown that the theory has seven physical plane 
waves. In addition, we investigate the gravitational radiation from binary 
systems in massive conformal gravity. We find that the theory with large 
graviton mass can reproduce the orbit of binaries by the emission of 
gravitational waves.  
\end{abstract}

\thispagestyle{empty}
\vfill
 \par
\bigskip
\noindent * felfrafar@hotmail.com \par
\end{titlepage}
\newpage

\section{Introduction}
\label{sec1}

Over the years several alternative theories of gravity have emerged in the 
attempt to solve some of the problems presented by the general theory of 
relativity, such as the dark matter and dark energy problems. Besides solving 
these problems, for an alternative theory of gravity to be considered 
consistent, it must also reproduce the successful predictions of general
relativity. One of these recently confirmed predictions is the existence of 
gravitational waves \cite{Ab1,Ab2,Ab3,Li}. 

Among the many alternative theories of gravity that have already studied 
the gravitational waves phenomenology is conformal gravity (CG). 
It was shown that the plane wave of this theory is composed of the usual 
plane wave of general relativity plus a plane wave that grows linearly in 
time \cite{Rie}, which causes the energy carried by the CG plane wave to 
diverges in momentum space \cite{RYang}. 

In this paper, we intend to study the behavior of gravitational waves in 
another alternative theory of gravity with conformal symmetry called massive 
conformal gravity (MCG) \cite{Faria1}. In Section \ref{sec2}, we present an 
introduction of the MCG theory. In Section \ref{sec3}, we find the plane wave 
solution of the linearized MCG field equations. In Section \ref{sec4}, we 
discuss the energy-momentum tensor of the MCG plane wave. In Section 
\ref{sec5}, we evaluate the radiated energy from a binary system in MCG.
Finally, in Section \ref{sec6}, we provide a brief conclusion about the 
results found in the paper. 
 

\section{Massive conformal gravity}
\label{sec2}


Let us consider the total MCG action\footnote{Here we consider units in which 
$c=\hbar=1$.} \cite{Faria2}
\begin{equation}
S_{\textrm{tot}} = \frac{1}{\kappa^2}\int{d^{4}x} \, \sqrt{-g}\bigg[\varphi^{2}
R + 6\partial^{\mu}\varphi\partial_{\mu}\varphi - \frac{1}{2m^2} 
C^{\alpha\beta\mu\nu}C_{\alpha\beta\mu\nu} \bigg] 
+ \int{d^{4}x}\mathcal{L}_{m}, 
\label{1}
\end{equation}
where $\kappa^{2} = 32\pi G/3$,  $\varphi$ is a scalar field called dilaton, 
$m$ is a constant with dimension of mass,
\begin{equation}
C^{\alpha\beta\mu\nu}C_{\alpha\beta\mu\nu} = R^{\alpha\beta\mu\nu}
R_{\alpha\beta\mu\nu} - 4R^{\mu\nu}R_{\mu\nu} + R^2
+ 2\left(R^{\mu\nu}R_{\mu\nu} - \frac{1}{3}R^{2}\right)
\label{2}
\end{equation}
is the Weyl tensor squared, $R^{\alpha}\,\!\!_{\mu\beta\nu} 
= \partial_{\beta}\Gamma^{\alpha}_{\mu\nu} + \cdots$ is the Riemann 
tensor, $R_{\mu\nu} = R^{\alpha}\,\!\!_{\mu\alpha\nu}$ is the Ricci tensor, 
$R = g^{\mu\nu}R_{\mu\nu}$ is the scalar curvature, and 
$\mathcal{L}_{m} = \mathcal{L}_{m}(g_{\mu\nu},\Psi)$ is the Lagrangian 
density of the matter field $\Psi$. 

The variation of the total action (\ref{1}) with respect to $g^{\mu\nu}$ 
and $\varphi$ gives the MCG field equations
\begin{equation}
\varphi^{2}G_{\mu\nu}
+  6 \partial_{\mu}\varphi\partial_{\nu}\varphi - 3g_{\mu\nu}\partial^{\rho}
\varphi\partial_{\rho}\varphi   + g_{\mu\nu} \nabla^{\rho}\nabla_{\rho} 
\varphi^{2} - \nabla_{\mu}\nabla_{\nu} \varphi^{2} 
- m^{-2}W_{\mu\nu} = \frac{1}{2}\kappa^2 T_{\mu\nu},
\label{3}
\end{equation}
\begin{equation}
\left(\nabla^{\rho}\nabla_{\rho} - \frac{1}{6}R\right)\varphi = 0,
\label{4}
\end{equation}
where
\begin{eqnarray}
W_{\mu\nu} &=& \nabla^{\rho}\nabla_{\rho}R_{\mu\nu} 
- \frac{1}{3}\nabla_{\mu}\nabla_{\nu}R  -\frac{1}{6}g_{\mu\nu}\nabla^{\rho}
\nabla_{\rho}R  + 2R^{\rho\sigma}R_{\mu\rho\nu\sigma} 
-\frac{1}{2}g_{\mu\nu}R^{\rho\sigma}R_{\rho\sigma}  \nonumber \\ &&
- \frac{2}{3}RR_{\mu\nu}  + \frac{1}{6}g_{\mu\nu}R^2
\label{5}
\end{eqnarray}
is the Bach tensor,
\begin{equation}
G_{\mu\nu} = R_{\mu\nu} - \frac{1}{2}g_{\mu\nu}R
\label{6}
\end{equation}
is the Einstein tensor, and
\begin{equation}
T_{\mu\nu} = \frac{2}{\sqrt{-g}} \frac{\delta \mathcal{L}_{m}}
{\delta g^{\mu\nu}}
\label{7}
\end{equation}
is the matter energy-momentum tensor.

Besides being invariant under coordinate 
transformations, the field equations (\ref{3}) and (\ref{4}) are 
also invariant under the conformal transformations
\begin{equation}
\tilde{\Phi} = \Omega(x)^{-\Delta_{\Phi}}\Phi,
\label{8}
\end{equation}
where $\Omega(x)$  is an arbitrary function of the spacetime coordinates, 
and $\Delta_{\Phi}$ is the scaling dimension of the field $\Phi$, whose 
values are $-2$ for the metric field, $0$ for gauge bosons, $1$ for 
scalar fields, and $3/2$ for fermions. Using the conformal invariance of 
the theory, we can impose the unitary gauge $\varphi = \varphi_{0} = 1$. 
In this case, the field equations (\ref{3}) and (\ref{4}) becomes
\begin{equation}
G_{\mu\nu} - m^{-2}W_{\mu\nu} = \frac{1}{2}\kappa^2 T_{\mu\nu},
\label{9}
\end{equation}
\begin{equation}
R = 0.
\label{10}
\end{equation}

Taking the trace of (\ref{9}) and comparing with (\ref{10}), we find that 
MCG couples only with matter whose energy-momentum tensor is traceless, which 
is a feature of conformally invariant theories. This is the 
case of the standard model of particle physics without a Higgs mass term, 
for instance. For simplicity, we consider the conformally invariant matter 
Lagrangian density \cite{Man}
\begin{equation}
\mathcal{L}_{m} = -\sqrt{-g}\Bigg[S^{2}R + 6\partial^{\mu}S\partial_{\mu}S 
+ \lambda S^{4} + \frac{i}{2}\left(\, \overline{\psi}
\gamma^{\mu}D_{\mu}\psi - D_{\mu}\overline{\psi}\gamma^{\mu}\psi \right) 
+ \mu S\overline{\psi}\psi\Bigg],
\label{11}
\end{equation}
where $S$ is a scalar Higgs field, $\lambda$ and $\mu$ are 
dimensionless coupling constants, $\overline{\psi} = \psi^{\dagger}
\gamma^{0}$ is the adjoint fermion field, $D_{\mu} = \partial_{\mu} 
+ [\gamma^{\nu},\partial_{\mu}\gamma_{\nu}]/8 - [\gamma^{\nu},\gamma_{\lambda}]
\Gamma^{\lambda}_{\mu\nu}/8$, and $\gamma^{\mu}$ 
are the general relativistic Dirac matrices, which satisfy the anticommutation 
relation $\{\gamma^{\mu},\gamma^{\nu}\} = 2g^{\mu\nu}$.

Before proceeding, it is worth noting that both the coordinate and the 
conformal symmetries of the theory 
allow the introduction of a quartic self-interaction term 
of the dilaton field in the gravitational part of the total MCG action (\ref{1}). 
The reason why we do not consider such a term is that its inclusion makes the 
flat metric no longer a solution of the vacuum field equations, which invalidates 
the usual $S$-matrix formulation. Additionally, we can include a coupling between 
the dilaton and the Higgs fields in the matter Lagrangian density (\ref{11}). 
However, we neglect this coupling because it leads to a nonvanishing trace of the matter 
energy-momentum tensor, as we can see by considering the variation of the modified 
$\mathcal{L}_{m}$ with respect to $\varphi$ on the right side of 
(\ref{4}) and comparing the resulting field equation with the trace of (\ref{3}).

The variation of (\ref{11}) with respect to $\overline{\psi}$ and $\psi$ gives 
the field equations
\begin{equation}
i\gamma^{\mu}D_{\mu}\psi + \mu S \psi = 0,
\label{12}
\end{equation}
\begin{equation}
iD_{\mu}\overline{\psi}\gamma^{\mu} - \mu S \overline{\psi} = 0.
\label{13}
\end{equation}
Substituting (\ref{11}) into (\ref{7}), and using (\ref{12}) and (\ref{13}), 
we obtain the matter energy-momentum tensor
\begin{eqnarray}
T_{\mu\nu} &=& 2g_{\mu\nu}\nabla^{\rho}S\nabla_{\rho}S
-8\nabla_{\mu}S\nabla_{\nu}S +4 S\nabla_{\mu}\nabla_{\nu} S  
- 4g_{\mu\nu}S\nabla^{\rho}\nabla_{\rho} S \nonumber \\ &&
+ 2S^{2}G_{\mu\nu} + T^{f}_{\mu\nu} - g_{\mu\nu}\lambda S^4,
\label{14}
\end{eqnarray}
where
\begin{equation}
T^{f}_{\mu\nu} = \frac{i}{4}\big(\, \overline{\psi}
\gamma_{\mu}D_{\nu}\psi - D_{\nu}\overline{\psi}\gamma_{\mu}\psi 
+ \overline{\psi}\gamma_{\nu}D_{\mu}\psi - D_{\mu}\overline{\psi}\gamma_{\nu}
\psi \big)
\label{15}
\end{equation}
is the fermion energy-momentum tensor.

Considering that, at scales below the electroweak scale, the Higgs field 
acquires a spontaneously broken constant vacumm expectation value $S_{0}$, 
we find that (\ref{14}) reduces to 
\begin{equation}
T_{\mu\nu} = 2S_{0}^{2}G_{\mu\nu} + T^{f}_{\mu\nu} - 
g_{\mu\nu}\lambda S_{0}^4.
\label{16}
\end{equation}
Taking the trace of (\ref{16}) and substituting into the trace of (\ref{9}), 
we obtain
\begin{equation}
-R = \frac{1}{2}\left( -2S_{0}^{2}R + T^{f} - 
4\lambda S_{0}^4  \right),
\label{17}
\end{equation}
where $T^{f} = g^{\mu\nu}T^{f}_{\mu\nu}$.
The additional use of (\ref{10}) then gives the relation
\begin{equation}
\lambda S_{0}^4  = \frac{1}{4}T^{f}.
\label{18}
\end{equation}
Finally, substituting this relation back into (\ref{16}), we arrive at
\begin{equation}
T_{\mu\nu} = 2S_{0}^{2}G_{\mu\nu} + T^{T}_{\mu\nu},
\label{19}
\end{equation}
where
\begin{equation}
T^{T}_{\mu\nu} = T^{f}_{\mu\nu} - \frac{1}{4}g_{\mu\nu}T^{f}
\label{20}
\end{equation}
is the traceless part of the fermion energy-momentum tensor.


\section{Plane gravitational waves}
\label{sec3}


In order to find the MCG gravitational wave equations, we must perturb the 
metric according to
\begin{equation}
g_{\mu\nu} = \eta_{\mu\nu} + h_{\mu\nu}, 
\label{21}
\end{equation}
where $\eta_{\mu\nu} = \mathrm{diag}( -1, +1, +1, +1)$ and 
$|h_{\mu\nu}| \ll 1$. Then, using (\ref{19}), we find that up to first 
order in the perturbation $h_{\mu\nu}$, the field equations (\ref{9}) 
and (\ref{10}) become\footnote{From now on the indexes $(i)$ indicate 
the order of the expansion in $h_{\mu\nu}$.}
\begin{equation}
G^{(1)}_{\mu\nu} 
- m^{-2}W^{(1)}_{\mu\nu} = \frac{1}{2}\kappa^2  T^{(0)}_{\mu\nu},
\label{22}
\end{equation}
\begin{equation}
R^{(1)} = 0,
\label{23}
\end{equation}
where
\begin{equation}
T^{(0)}_{\mu\nu} = 2S_{0}^{2}G^{(0)}_{\mu\nu} + T^{T(0)}_{\mu\nu} 
= T^{f(0)}_{\mu\nu} - \frac{1}{4}\eta_{\mu\nu}T^{f(0)}
\label{24}
\end{equation}
is the zero-order matter energy-momentum tensor,
\begin{equation}
W^{(1)}_{\mu\nu} = \Box R^{(1)}_{\mu\nu} 
- \frac{1}{3}\partial_{\mu}\partial_{\nu}R^{(1)}  -\frac{1}{6}\eta_{\mu\nu}
\Box R^{(1)}
\label{25}
\end{equation}
is the first-order Bach tensor,
\begin{equation}
G^{(1)}_{\mu\nu} = R^{(1)}_{\mu\nu} - \frac{1}{2}\eta_{\mu\nu}R^{(1)}
\label{26}
\end{equation}
is the first-order Einstein tensor,
\begin{equation}
R^{(1)}_{\mu\nu} = \frac{1}{2} \left( \partial_{\mu}\partial^{\rho}
h_{\rho\nu} + \partial_{\nu}\partial^{\rho}h_{\rho\mu} 
- \Box h_{\mu\nu} 
- \partial_{\mu}\partial_{\nu}h  \right)
\label{27}
\end{equation}
is the first-order Ricci tensor, and
\begin{equation}
R^{(1)} =  \partial^{\mu}\partial^{\nu}h_{\mu\nu} 
- \Box h
\label{28}
\end{equation} 
is the first-order scalar curvature, with $\Box = \partial^{\mu}\partial_{\mu}$ 
and $h = \eta^{\mu\nu}h_{\mu\nu}$. 

It is not difficult to see that both (\ref{22}) and (\ref{23}) are invariant 
under the coordinate gauge transformation 
\begin{equation}
h'_{\mu\nu} = h_{\mu\nu} + \partial_{\mu}\xi_{\nu} + \partial_{\nu}\xi_{\mu},
\label{29}
\end{equation}
where $\xi_{\mu}$ is an arbitrary spacetime dependent vector field. 
By imposing the gauge condition
\begin{equation}
\partial^{\mu}h_{\mu\nu} - \frac{1}{2}\partial_{\nu}h = 0,
\label{30}
\end{equation}
which fix the coordinate gauge freedom up to residual gauge parameter 
satisfying the subsidiary condition
\begin{equation}
\Box \xi_{\mu} = 0,
\label{31}
\end{equation}
and using (\ref{23}), we find that (\ref{22}) reduces to
\begin{equation}
\left(m^{-2}\Box - 1 \right)\Box h_{\mu\nu} = \kappa^2 
T^{(0)}_{\mu\nu}.
\label{32}
\end{equation}

The  simplest physical solution to (\ref{32}) in vacuum ($T^{(0)}_{\mu\nu}=0$) 
is the plane wave
\begin{equation}
h_{\mu\nu} = a_{\mu\nu}\cos(k_{\rho}x^{\rho})  
+ b_{\mu\nu}\cos(q_{\rho}x^{\rho}),
\label{33}
\end{equation}
where $a_{\mu\nu}$ and $b_{\mu\nu}$ are symmetric wave polarization tensors, 
and $k_{\mu}$ and $q_{\mu}$ are wave vectors, 
which satisfies 
\begin{equation}
k^{\rho}k_{\rho} = 0, \ \ \ \ \ \  q^{\rho}q_{\rho} = -m^2.
\label{34}
\end{equation}
By substituting (\ref{33}) into (\ref{23}) and (\ref{30}), and using (\ref{34}), 
we obtain
\begin{equation}
k^{\mu}a_{\mu\nu} = \frac{1}{2}k_{\nu}a, \ \ \ \ \ \ q^{\mu}b_{\mu\nu} 
= 0, \ \ \ \ \ \ b = 0,
\label{35}
\end{equation}
where $a = \eta^{\mu\nu}a_{\mu\nu}$ and $b = \eta^{\mu\nu}b_{\mu\nu}$.

The symmetry of the polarization tensors $a_{\mu\nu}$ and 
$b_{\mu\nu}$ means that each of them has ten independent components. The 
conditions (\ref{35}) reduce the independent components of $a_{\mu\nu}$ 
to six and of $b_{\mu\nu}$ to five. In addition, we can choose a solution of 
(\ref{31}) to impose four more conditions on $a_{\mu\nu}$. For instance, 
choosing
\begin{equation}
\xi_{\mu} =  \epsilon_{\mu}\sin(k_{\rho}x^{\rho}) ,
\label{36}
\end{equation}
and substituting into (\ref{29}) together with (\ref{33}), we arrive at
\begin{equation}
a'_{\mu\nu} = a_{\mu\nu} + k_{\mu}\epsilon_{\nu} + k_{\nu}\epsilon_{\mu}.
\label{37}
\end{equation}
Since $\epsilon_{\mu}$ is arbitrary, we can select it to
impose four more conditions on $a_{\mu\nu}$. In particular, we can choose 
$\epsilon_{\mu}$ such that
\begin{equation}
 a_{0i} = 0, \ \ \ \ \ \ \delta^{ij}a_{ij} = 0,
\label{38} 
\end{equation}
which reduce the independent components of  $a_{\mu\nu}$ to just two. Thus, 
we conclude that the MCG plane gravitational wave (\ref{33}) has seven 
propagating degrees of freedom, represented by the two independent components 
of $a_{\mu\nu}$ and the five independent components of $b_{\mu\nu}$, which is 
consistent with recent results obtained in the literature \cite{Faria2,Hols}.


\section{Gravitational energy-momentum tensor}
\label{sec4}


We can see from (\ref{32}) that the total linearized MCG Lagrangian density 
is dynamically equivalent to
\begin{equation}
\mathcal{L}_{\textrm{tot}} = -\frac{1}{4\kappa^2}\left(m^{-2}\Box 
h^{\mu\nu}\Box h_{\mu\nu} + \partial^{\rho}h^{\mu\nu}\partial_{\rho}h_{\mu\nu} 
\right) + \frac{1}{2}h^{\mu\nu}T^{(0)}_{\mu\nu}.
\label{39}
\end{equation}
Inserting the gravitational part of (\ref{39}) into the canonical 
energy-momentum tensor \cite{Capo}
\begin{equation}
t^{\mu\nu} =   \bigg\langle \left[ \partial_{\rho}\frac{\partial 
\mathcal{L}}{\partial (\partial_{\mu}\partial_{\rho}
h_{\alpha\beta})} - \frac{\partial \mathcal{L}}
{\partial (\partial_{\mu}h_{\alpha\beta})}
\right]\partial^{\nu}h_{\alpha\beta} - \frac{\partial 
\mathcal{L}}{\partial (\partial_{\mu}\partial_{\rho}
h_{\alpha\beta})}\partial^{\nu}\partial_{\rho}h_{\alpha\beta} 
+ \eta^{\mu\nu}\mathcal{L} \bigg\rangle,
\label{40}
\end{equation}
integrating by parts, and using (\ref{32}) in vacuum, we 
obtain
\begin{equation}
t_{\mu\nu} = \frac{1}{2\kappa^2 }\left\langle  
2m^{-2}\partial_{\mu}\partial_{\nu}h^{\alpha\beta}\Box 
h_{\alpha\beta} + \partial_{\mu}h^{\alpha\beta} 
\partial_{\nu}h_{\alpha\beta} \right\rangle,
\label{41}
\end{equation}
where the angle brackets denote the average over a macroscopic region.

The substitution of the plane wave (\ref{33}) into (\ref{41}) gives 
\begin{equation}
t_{\mu\nu} = \frac{1}{2\kappa^2}
\left[\left(a^{\alpha\beta}a_{\alpha\beta}\right)k_{\mu}k_{\nu}
- \left(b^{\alpha\beta}b_{\alpha\beta}\right)q_{\mu}q_{\nu} \right],
\label{42}
\end{equation}
where we used (\ref{34}). We can see from (\ref{42}) that the energies of 
the waves with the two physical polarizations $a_{\mu\nu}$ are positive, 
while those of the ones with the five physical polarizations $b_{\mu\nu}$ 
are negative. Since these waves do not interact with each other, energy 
cannot flow between them so that there is no violation of energy conservation.


\section{Gravitational waves from a binary system}
\label{sec5}


In order to analyze the gravitational waves created from binary systems, we
need to solve the fourth-order differential equation (\ref{32}) in the presence 
of matter ($T^{(0)}_{\mu\nu} \neq 0$). For simplicity, we can split 
$h_{\mu\nu}$ according to
\begin{equation}
h_{\mu\nu} = A_{\mu\nu} + B_{\mu\nu},
\label{43}
\end{equation}
where $A_{\mu\nu}$ and $B_{\mu\nu}$ obey the second-order differential 
equations
\begin{equation}
\Box A_{\mu\nu} = -\kappa^2 T^{(0)}_{\mu\nu},
\label{44}
\end{equation}
\begin{equation}
\left(\Box - m^{2}\right) B_{\mu\nu} = \kappa^2  T^{(0)}_{\mu\nu}.
\label{45}
\end{equation}

In the frequency domain, the field equations (\ref{44}) and (\ref{45}) 
become
\begin{equation}
\left( \nabla^{2} + \omega^2 \right) \tilde{A}_{\mu\nu}(\omega,\textbf{x}) 
= -\kappa^2  \tilde{T}^{(0)}_{\mu\nu}(\omega,\textbf{x}),
\label{46}
\end{equation}
\begin{equation}
\left( \nabla^{2} + \omega^2 - m^{2}\right) \tilde{B}_{\mu\nu}(\omega,
\textbf{x})
 = \kappa^2 \tilde{T}^{(0)}_{\mu\nu}(\omega,\textbf{x}),
\label{47}
\end{equation}
where $\nabla^{2}$ is the Laplacian, $\omega$ is the frequency of the wave, 
and the tilde denotes the Fourier transform. The general solutions to 
(\ref{46}) and (\ref{47}) are given by
\begin{equation}
\tilde{A}_{\mu\nu}(\omega, \textbf{x}) = - \kappa^2 \int{
d^{3}\textbf{x}'}\tilde{G}_{A}(\omega,\textbf{r})\tilde{T}^{(0)}_{\mu\nu}
(\omega, \textbf{x}'),
\label{48}
\end{equation}
\begin{equation}
\tilde{B}_{\mu\nu}(\omega, \textbf{x}) = \kappa^2 \int{
d^{3}\textbf{x}'}\tilde{G}_{B}(\omega,\textbf{r})\tilde{T}^{(0)}_{\mu\nu}
(\omega, \textbf{x}'),
\label{49}
\end{equation}
where the frequency domain Green functions $\tilde{G}_{A}(\omega,\textbf{r})$ 
and $\tilde{G}_{B}(\omega,\textbf{r})$ are defined by
\begin{equation}
 \left( \nabla^{2} + \omega^2 \right) \tilde{G}_{A}(\omega,\textbf{r})
= \sqrt{8\pi}\delta^{(3)}(\textbf{r}),
\label{50}
\end{equation}
\begin{equation}
\left( \nabla^{2} + \omega^2 - m^{2}\right) \tilde{G}_{B}(\omega,\textbf{r})
= \sqrt{8\pi}\delta^{(3)}(\textbf{r}),
\label{51}
\end{equation}
with $\textbf{r} = \textbf{x} - \textbf{x}'$ being the difference between the 
positions of the observer ($\textbf{x}$) and the source ($\textbf{x}'$).

It follows from (\ref{50}) that
\begin{equation}
\tilde{G}_{A}(\omega,\textbf{r}) = -\frac{ e^{i\omega|\textbf{r}|}}
{{4\pi|\textbf{r}|}},
\label{52}
\end{equation}
and from (\ref{51}) that
\begin{equation}
\tilde{G}_{B}(\omega,\textbf{r}) = -\frac{e^{ik_{\omega}|\textbf{r}|}
\Theta(\omega - m)+ \mbox{c.c.}\Theta(-\omega - m)}
{4\pi|\textbf{r}|}
\label{53}
\end{equation}
for $m^2 < \omega^2$, and
\begin{equation}
\tilde{G}_{B}(\omega,\textbf{r}) = -\frac{e^{-k_{m}|\textbf{r}|}
\Theta(m -|\omega|)}{4\pi|\textbf{r}|}
\label{54}
\end{equation}
for $m^2 > \omega^2$, where c.c. is the complex conjugate
of the exponential function,  $\Theta$ is the Heaviside step function, 
$k_{\omega} = \sqrt{\omega^2-m^2}$, and $k_{m} = \sqrt{m^2-\omega^2}$.

Substituting (\ref{52}) into (\ref{48}), transforming back to real space, 
and using the far zone approximation ($|\textbf{x}| \approx 
|\textbf{x} - \textbf{x}'|$), we can write the spatial components of 
$A_{\mu\nu}$ in the form
\begin{equation}
A_{ij}(t,\textbf{r}) = \frac{\kappa^2}{8\pi r}
\left[\int_{-\infty}^{+\infty}\frac{d\omega}{2\pi}e^{-i\omega(t-r)}\right]
\omega^{2}\tilde{Q}_{ij}(\omega),
\label{55}
\end{equation}
where $r = |\textbf{x}|$ is the distance between the observer and the 
source, and $\tilde{Q}_{ij}(\omega)$ is the Fourier 
transform of the reduced quadrupole moment
\begin{equation}
Q_{ij}(t) = \int{d^{3}\textbf{r}}\left(x_{i}x_{j}
- \frac{1}{3}\delta_{ij}r^2\right)T^{(0)}_{00}(t, \textbf{r}).
\label{56}
\end{equation}
Following the same steps for $B_{\mu\nu}$, but now with the substitution 
of (\ref{53}) and (\ref{54}) into (\ref{49}) , we obtain
\begin{equation}
B_{ij}(t,\textbf{r}) = -\frac{\kappa^2}{8\pi r}
\bigg[\int_{m}^{\infty}\frac{d\omega}{2\pi}e^{-i\omega t}e^{ik_{\omega}r}
+\int_{-\infty}^{-m}\frac{d\omega}{2\pi}e^{-i\omega t}e^{-ik_{\omega}r}\bigg]
\omega^{2}\tilde{Q}_{ij}(\omega)
\label{57}
\end{equation}
for $m^2 < \omega^2$, and
\begin{equation}
B_{ij}(t,\textbf{r}) = -\frac{\kappa^2}{8\pi r}\left[
\int_{-\infty}^{m}\frac{d\omega}{2\pi}e^{-i\omega t}e^{-k_{m}r}\right]
\omega^{2}\tilde{Q}_{ij}(\omega)
\label{58}
\end{equation}
for $m^2 > \omega^2$.

In the case of a circular binary system formed by a pair of masses 
$m_{1}$ and $m_{2}$, separated by a distance $d$, orbiting each other in 
the $xy$-plane with frequency $\omega_{s} = \omega/2$, we have 
$T^{f(0)}_{\mu\nu}(t,\textbf{x}) = \mu \delta^{0}_{\mu}\delta^{0}_{\nu}
\delta^3(\textbf{x})$. The substitution of this value into (\ref{24}) gives
\begin{equation}
T^{(0)}_{00}(t,\textbf{r}) = \frac{3}{4}\mu \delta^3(\textbf{r}),
\label{59}
\end{equation}
where $\mu = m_{1}m_{2}/(m_{1}+m_{2})$ is the reduced mass. By inserting 
(\ref{59}) and the relative coordinates
\begin{equation}
x_{1} = -d\sin(\omega_{s}t), \ \ \ \ \ x_{2} = d\cos(\omega_{s}t), \ \ \ \ \ 
x_{3} = 0,
\label{60}
\end{equation}
into (\ref{56}), and taking the Fourier transform, we find
\begin{equation}
\tilde{Q}_{11}(\omega) = \frac{3\pi\mu d^{2}}{8}\left[\delta(\omega) 
- \delta(\omega + 2\omega_{s}) - \delta(\omega - 2\omega_{s})\right],
\label{61}
\end{equation}
\begin{equation}
\tilde{Q}_{22}(\omega) = \frac{3\pi\mu d^{2}}{8}\left[\delta(\omega) 
+ \delta(\omega + 2\omega_{s}) + \delta(\omega - 2\omega_{s})\right],
\label{62}
\end{equation}
\begin{equation}
\tilde{Q}_{12}(\omega) = \frac{3\pi\mu d^{2}}{8i}\left[\delta(\omega 
- 2\omega_{s}) - \delta(\omega + 2\omega_{s})\right],
\label{63}
\end{equation}
where we omitted the term proportional to $\delta_{ij}$ in (\ref{56}) 
because it does not contribute to the radiated energy.

The insertion of (\ref{61})-(\ref{63}) into (\ref{55}) gives
\begin{equation}
A_{11}(t,r) = - A_{22}(t,r) = \frac{2\mu d^{2}\omega_{s}^2}{r} 
\cos{(2\omega_{s}t_{\textrm{ret}})},
\label{64}
\end{equation}
\begin{equation}
A_{12}(t,r) = A_{21}(t,r) = \frac{2\mu d^{2}\omega_{s}^2}{r}
\sin{(2\omega_{s}t_{\textrm{ret}})},
\label{65}
\end{equation}
where $t_{\textrm{ret}} = t -r$ is the retarded time. In the same way, 
substituting (\ref{61})-(\ref{63}) into (\ref{57}) and (\ref{58}), we 
arrive at 
\begin{equation}
B_{11}(t,r) = - B_{22}(t,r) = -\frac{2\mu d^{2}\omega_{s}^2}{r}
\cos{(2\omega_{s}t_{m})},
\label{66}
\end{equation}
\begin{equation}
B_{12}(t,r) = B_{21}(t,r) = -\frac{2\mu d^{2}\omega_{s}^2}{r}
\sin{(2\omega_{s}t_{m})},
\label{67}
\end{equation}
for $m^2 < 4\omega_{s}^2$, and
\begin{equation}
B_{11}(t,r) = - B_{22}(t,r) = -\frac{2\mu d^{2}\omega_{s}^2}{r}
e^{-k_{m}r}\cos{(2\omega_{s}t)},
\label{68}
\end{equation}
\begin{equation}
B_{12}(t,r) = B_{21}(t,r) = -\frac{2\mu d^{2}\omega_{s}^2}{r}
e^{-k_{m}r}\sin{(2\omega_{s}t)},
\label{69}
\end{equation}
for $m^2 > 4\omega_{s}^2$, where $t_{m}= t - v_{m}r$ is the travel time, with 
$v_{m}=\sqrt{1-m^{2}/(4\omega_{s}^2)}$ being the speed of the massive 
gravitational wave.

The rate of energy loss from a source, in the far field limit, is given by
\begin{equation}
\dot{E} = - r^2\int_{\partial V}{d\Omega} \,  t^{0i}n_{i},
\label{70}
\end{equation}
where the dot is the derivative with respect to time, 
$d\Omega = \sin\theta d\theta d\phi$ is the differential solid angle, 
$\partial V$ is the surface of a spherical shell with volume $V$ centered 
around the source, and $n_{i}$ are the components of the 
spatial unit vector pointing from the source to the observer. 

By substituting (\ref{43}) into (\ref{41}), integrating by parts, 
and using (\ref{44}) and (\ref{45}) in vacuum, we obtain
\begin{equation}
t^{0i}n_{i} = \frac{1}{2\kappa^2}n_{i}\big\langle \partial^{0}
A^{\alpha\beta}\partial^{i} A_{\alpha\beta} - \partial^{0}B^{\alpha\beta}
\partial^{i} B_{\alpha\beta}   \big\rangle. 
\label{71}
\end{equation}
We can see from (\ref{64}) and (\ref{65}) that
\begin{equation}
\partial_{i}A_{\alpha\beta} \approx - n_{i}\partial_{0}A_{\alpha\beta},
\label{72}
\end{equation}
and from (\ref{66})-(\ref{69}) that
\begin{equation}
\partial_{i}B_{\alpha\beta} \approx - n_{i}v_{m}\partial_{0}B_{\alpha\beta}
\label{73}
\end{equation}
for $m^2 < 4\omega_{s}^2$, and
\begin{equation}
\partial_{i}B_{\alpha\beta} \approx - n_{i}k_{m}B_{\alpha\beta}
\label{74}
\end{equation}
for $m^2 > 4\omega_{s}^2$, where we neglected terms of order $1/r^2$. 
In addition, considering the conservation and the traceless condition of 
$T^{(0)}_{\mu\nu}$, and the traceless condition of $Q_{ij}$, it follows 
from (\ref{48}), (\ref{49}), (\ref{55}), (\ref{57}) and (\ref{58}) that 
$A_{\mu\nu}$ and $B_{\mu\nu}$ obey the traceless-transverse conditions
\begin{equation}
\partial^{i}A_{ij} = 0, \ \ \ \ \ \  A_{0i} = 0, \ \ \ \ \ \ 
\delta^{ij}A_{ij} = 0,
\label{75}
\end{equation}
\begin{equation}
\partial^{i}B_{ij} = 0, \ \ \ \ \ \  B_{0i} = 0, \ \ \ \ \ \ 
\delta^{ij}B_{ij} = 0.
\label{76}
\end{equation}

By using the combination of (\ref{72})-(\ref{76}), and $n^{i}n_{i} = 1$, 
we can write (\ref{71}) as
\begin{equation}
t^{0i}n_{i} \approx \frac{1}{2\kappa^2}\Lambda_{ijkl}\big\langle  
\partial_{0}A^{ij}\partial_{0} A^{kl} 
- v_{m}\partial_{0}B^{ij}\partial_{0} B^{kl} \big\rangle 
\label{77}
\end{equation}
for  $m^2 < 4\omega_{s}^2$, and
\begin{equation}
t^{0i}n_{i} \approx \frac{1}{2\kappa^2}\Lambda_{ijkl}\big\langle  
\partial_{0}A^{ij}\partial_{0} A^{kl} 
- k_{m}B^{ij}\partial_{0} B^{kl} \big\rangle 
\label{78}
\end{equation}
for  $m^2 > 4\omega_{s}^2$, where
\begin{equation}
\Lambda_{ijkl} = P_{ik}P_{jl} -\frac{1}{2}P_{ij}P_{kl} 
\label{79}
\end{equation}
is the Lambda tensor, with
\begin{equation}
P_{ij} = \delta_{ij} - n_{i}n_{j}
\label{80} 
\end{equation}
being the traceless-transverse projection operator.

Inserting (\ref{77}) and (\ref{78}) into (\ref{70}), and using the surface 
integral
\begin{equation}
\int_{\partial V}{d\Omega}\Lambda_{ijkl} = \frac{2\pi}{15}\left( 
11\delta_{ik}\delta_{jl} -4 \delta_{ij}\delta_{kl} + \delta_{il}\delta_{jk} 
\right),
\label{81}
\end{equation}
we find
\begin{equation}
\dot{E} \approx - \frac{3 r^2}{40}\big\langle  
\partial_{0}A^{ij}\partial_{0} A_{ij}
-  v_{m}\partial_{0}B^{ij}
\partial_{0} B_{ij}
\big\rangle 
\label{82}
\end{equation}
for $m^2 < 4\omega_{s}^2$, and 
\begin{equation}
\dot{E} \approx - \frac{3 r^2}{40}\big\langle  
\partial_{0}A^{ij}\partial_{0} A_{ij} 
- k_{m}B^{ij}\partial_{0} B_{ij}
\big\rangle 
\label{83}
\end{equation}
for $m^2 > 4\omega_{s}^2$. 

Finally, substituting (\ref{64})-(\ref{69}) into 
(\ref{82}) and (\ref{83}), we arrive at
\begin{equation}
\dot{E} \approx \frac{3}{8}
\left(1-\sqrt{1-\frac{m^{2}}{4\omega_{s}^2}}
\,\right)\dot{E}_{\textrm{GR}}
\label{84}
\end{equation}
for $m^2 < 4\omega_{s}^2$, and 
\begin{equation}
\dot{E} \approx \frac{3}{8}\,
\dot{E}_{\textrm{GR}}
\label{85}
\end{equation}
for $m^2 > 4\omega_{s}^2$, where
\begin{equation}
\dot{E}_{\textrm{GR}} = -\frac{32G\mu^{2}d^{4}\omega_{s}^6}{5}
\label{86}
\end{equation}
is the standard  energy loss of general relativity.

Experiments on the inverse square law of the MCG gravitational 
potential of a point particle with a mass $M$, 
which is given by \cite{Faria2} 
\begin{equation}
\phi(r) = -\frac{GM}{r}\left( 1 - e^{-mr}\right),
\label{87}
\end{equation}
constrain the graviton mass to the ranges $m< 10^{-22}\, \textrm{eV}$ for 
$m^2 < 4\omega_{s}^2$ and $m> 10^{-2}\, \textrm{eV}$ for $m^2 > 4\omega_{s}^2$ 
\cite{Adel}. In particular, considering the orbital frequency 
$\omega_{s} \approx 1.3\times 10^{-20}\, \textrm{eV}$ of the binary 
system PSR J1012 + 5307 formed by a neutron star and a white dwarf in 
quasi-circular motion \cite{Laz,Nic,Cal}, we have that $m^2/4\omega_{s}^2 < 
10^{-5}$ for $m^2 < 4\omega_{s}^2$. By substituting this value into (\ref{84}), 
we find
\begin{equation} 
\dot{E} \lessapprox 10^{-6} \dot{E}_{\textrm{GR}}
\label{88}
\end{equation}
for $m^2 < 4\omega_{s}^2$.

The energy loss of the binary system results in a decay of its orbital period 
$P$, which can be written as
\begin{equation}
\frac{\dot{P}}{P} = \frac{\dot{d}}{2d} - \frac{\dot{\phi}'}{2\phi'},
\label{89}
\end{equation}
where $\phi'(d)=\mu^{-1}\partial_{d}U(d)$  is the derivative of the 
gravitational potential $\phi$ with respect to $d$, with $U$ being the 
gravitational potential energy. It follows from (\ref{87}) that
\begin{equation}
U(d) = -\frac{Gm_{1}m_{2}}{d}\left( 1 - e^{-md}\right).
\label{90}
\end{equation}
Substituting this result into (\ref{89}), we find
\begin{equation}
\frac{\dot{P}}{P} \approx -\frac{3}{2}\frac{|\dot{E}|}{|E_{\textrm{GR}}|}
\label{91}
\end{equation}
for both $m^2 < 4\omega_{s}^2$ and $m^2 > 4\omega_{s}^2$, where 
$|E_{\textrm{GR}}| = G m_{1}m_{2}/2d$ and we considered $m^2d^2 \ll 1$ for 
$m^2 < 4\omega_{s}^2$.

The insertion of (\ref{88}) and (\ref{85}) into (\ref{91}) then gives
\begin{equation}
\frac{\dot{P}}{P} \lessapprox 10^{-6}\frac{\dot{P}_{\textrm{GR}}}
{P_{\textrm{GR}}}
\label{92}
\end{equation}
for $m^2 < 4\omega_{s}^2$, and
\begin{equation}
\frac{\dot{P}}{P} \approx \frac{3}{8}\frac{\dot{P}_{\textrm{GR}}}
{P_{\textrm{GR}}}
\label{93}
\end{equation}
for $m^2 > 4\omega_{s}^2$, where $\dot{P}_{\textrm{GR}}/P_{\textrm{GR}} 
= -3|\dot{E}_{\textrm{GR}}|/2|E_{\textrm{GR}}|$. We can see from (\ref{92}) 
that the MCG decay of the orbital period for $m^2 < 4\omega_{s}^2$ is several 
orders of magnitude smaller than in general relativity, which rules out 
the theory with a small graviton mass ($m< 10^{-22}\, \textrm{eV}$). On the 
other hand, the discrepancy seen in (\ref{93}) between the MCG decay of 
the orbital period for $m^2 > 4\omega_{s}^2$ and the general relativity result 
disappears if we include the first-order term 
$2S_{0}^{2}G^{(1)}_{\mu\nu}$ in the matter energy-momentum tensor (\ref{24}) 
as done in Ref. \cite{Capri} to CG\footnote{It is worth noting that CG 
with $\epsilon = +1$ considered in Ref. \cite{Capri} is different from MCG 
because the condition $R = 0$ does not hold in it \cite{Faria3}. Although this 
condition is irrelevant for the gravitational wave solutions from a binary 
system, it should leads to different results in other solutions such as the 
cosmological ones.}. Taking this into account, the theory with a large graviton 
mass ($m> 10^{-2}\, \textrm{eV}$) can explain the decrease of the orbital 
period of binary systems.


\section{Final remarks}
\label{sec6}


Here we have shown that the MCG plane wave has seven propagating 
degrees of freedom, two of which are massless and carry positive energies and 
the other five are massive and carry negative energies. Despite the presence 
of the waves with negative energies, the lack of interaction between them and 
the waves with positive energies means that the energies of all the seven MCG 
plane waves do not diverge.

The study on the radiated energy from a binary system restricts the MCG to 
large graviton mass, which give rise to a modification of general relativity 
only at high energies and small distances. Although this modification is the 
cause of MCG being renormalizable \cite{Faria4,Faria5}, it makes the theory 
unable to explain galaxy rotation curves without dark matter. However, the 
conformal symmetry of the matter part of MCG allows us to consider that the
Higgs mass is generated by the symmetry breaking of an extra scalar
field, which may be a good candidate for dark 
matter. Further studies are needed to figure this out.



\begin{thebibliography}{99}

\bibitem{Ab1}
B.P. Abbott et al., Phys. Rev. Lett. \textbf{116},  061102 (2016).

\bibitem{Ab2}
B.P. Abbott et al., Phys. Rev. Lett. \textbf{119}, 141101 (2017).

\bibitem{Ab3}
B.P. Abbott et al., Astrophys. J. \textbf{848}, L12 (2017).

\bibitem{Li}
T. Li et al., Sci. China Phys. Mech. Astron. \textbf{61}, 031011 (2018).

\bibitem{Rie}
R. J. Riegert, Phys. Lett. A \textbf{105}, 110 (1984).

\bibitem{RYang}
R. Yang, Phys. Lett. B \textbf{784}, 212 (2018).

\bibitem{Faria1}
F. F. Faria, Adv. High Energy Phys. \textbf{2014}, 520259 (2014).

\bibitem{Faria2}
F. F. Faria, Adv. High Energy Phys. \textbf{2019}, 7013012 (2019).

\bibitem{Hols}
P. H\"olscher, Phys. Rev. D \textbf{99}, 064039 (2019).

\bibitem{Man}
P.D. Mannheim, Gen. Relativ. Gravit. \textbf{22}, 289 (1990).

\bibitem{Capo}
S. Capozziello, M. Capriolo and M. Transirico, Ann. Phys. \textbf{525}, 1600376 (2017).

\bibitem{Adel}
E. Adelberger, J. Gundlach, B. Heckel, S. Hoedl, and S. Schlamminger, 
Prog. Part. Nucl. Phys. \textbf{62}, 102 (2009).

\bibitem{Laz}
K. Lazaridis et al., Mon. Not. R. Astron. Soc. \textbf{400}, 805 (2009).

\bibitem{Nic}
L. Nicastro, A. Lyne, D. Lorimer, P. Harrison, M. Bailes, and B. Skidmore, 
Mon. Not. R. Astron. Soc. \textbf{273}, L68 (1995).

\bibitem{Cal}
P.J. Callanan, P.M. Garnavich, and D. Koester, Mon. Not.
R. Astron. Soc. \textbf{298}, 207 (1998).

\bibitem{Capri}
 C. Caprini, P. H\"olscher, D.J. Schwarz, Phys. Rev. D \textbf{98}, 
084002 (2018).

\bibitem{Faria3}
F. F. Faria, Phys. Rev. D \textbf{99}, 048501 (2019).

\bibitem{Faria4}
F. F. Faria, Eur. Phys. J. C \textbf{76}, 188 (2016).

\bibitem{Faria5}
F. F. Faria, Eur. Phys. J. C \textbf{77}, 11 (2017).

\end{thebibliography}
\end{document}